\documentclass[aps,twocolumn]{revtex4}
\usepackage{MnSymbol}
\usepackage{graphicx}

\begin{document}

\title{Multiple-length-scale elastic instability mimics parametric resonance of nonlinear oscillators}
\author{Fabian Brau$^1$, Hugues Vandeparre$^1$, Abbas Sabbah$^1$, Christophe Poulard$^1$, Arezki Boudaoud$^{2}$ and Pascal Damman$^{1}$}
\affiliation{$^1$Laboratoire Interfaces $\&$ Fluides Complexes, CIRMAP, Universit\'e de Mons - UMONS, 20 Place du Parc, B-7000 Mons, Belgium}
\affiliation{$^2$Laboratoire de Physique Statistique, Ecole Normale Sup\'erieure, UPMC Paris 06, Universit\'e Paris Diderot, CNRS, 24 rue Lhomond, 75005 Paris, France} 

\date{\today}
\maketitle

{\bf Spatially confined rigid membranes reorganize their morphology in response to the imposed constraints. A crumpled elastic sheet presents a complex pattern of random folds focusing the deformation energy~\cite{witt07} while compressing a membrane resting on a soft foundation creates a regular pattern of sinusoidal wrinkles with a broad distribution of energy~\cite{bowd98,cerd03,vandeparre07,vandeparre08,huan07,jian07,poci08}. Here, we study the energy distribution for highly confined membranes and show the emergence of a new morphological instability triggered by a period-doubling bifurcation. A periodic self-organized focalization of the deformation energy is observed provided an up-down symmetry breaking, induced by the intrinsic nonlinearity of the elasticity equations, occurs. The physical model, exhibiting an analogy with parametric resonance in nonlinear oscillator, is a new theoretical toolkit to understand the morphology of various confined systems, such as coated materials or living tissues, e.g., wrinkled skin \cite{cerd03}, internal structure of lungs \cite{diam01}, internal elastica of an artery \cite{strupler}, brain convolutions \cite{rich75,toro05} or formation of fingerprints \cite{kuck04}. Moreover, it opens the way to new kind of microfabrication design of multiperiodic or chaotic (aperiodic) surface topography via self-organization.} 


Several theoretical approaches have been proposed to describe the wrinkling instability for very small compression ratio, {\it i.e.} near the instability threshold \cite{bowd98,jian07,cerd03}. However, the large compression domain remains largely unexplored with the notable exception of the wrinkle to fold transition observed in Ref.~\cite{poci08} for elastic membrane on liquid and the self-similar wrinkling patterns in skins \cite{efim04}. In the former case, the deformation of the membrane is progressively focalized into a single fold, concentrating all the bending energy. In contrast, for thin rigid membranes on elastomers, large compression induces perturbations of the initial wrinkles but the elasticity of the soft foundation maintains a regular periodic pattern whose complexity increases with the compression ratio. 

\begin{figure*}[!hbtp]
\centerline{\includegraphics[width=18.3cm,clip]{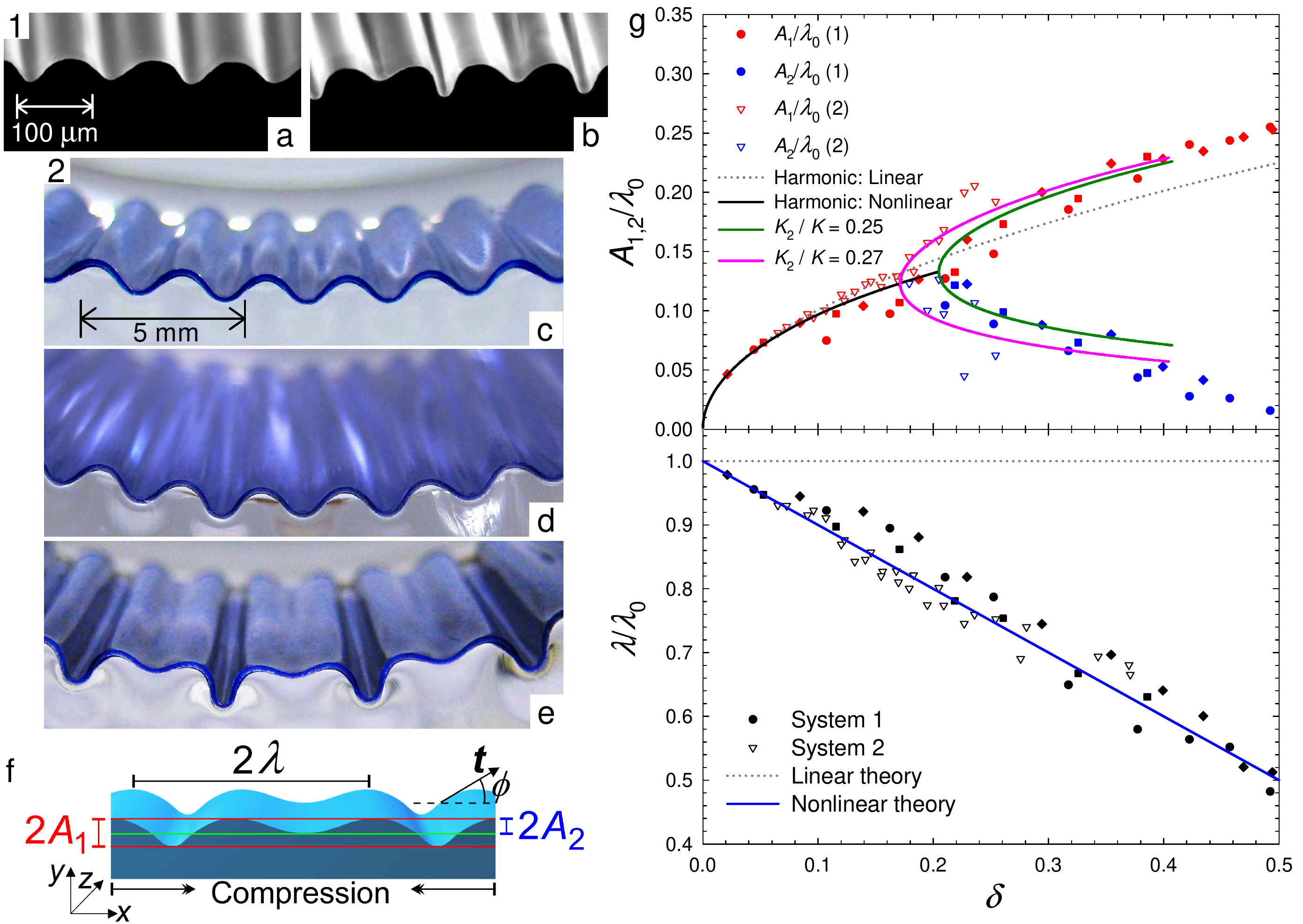}}
\caption{{\bf Evolution of morphologies, wavelengths and amplitudes with compression}. System 1 (a,b): A PDMS foundation is cured with UV/ozone which modifies the elastic properties of its surface. The thickness of the membrane is about $10-20$ $\mu$m depending on the irradiation time. The wavelength, $\lambda_0$, of the initial wrinkling instability is about $50-100$ $\mu$m. System 2 (c,d,e): A thin coloured stiff PDMS film is bound to a thick soft PDMS foundation. The thickness of the membrane is about 200 $\mu$m and the initial wrinkle wavelength is about $3$ mm. The compression ratios, $\delta$, are equal to $0.165$, $0.19$ and $0.24$ for panels c, d and e respectively. (f) The systems are compressed uniaxially along the $x$-axis. The wavelength and amplitudes of the wrinkles are measured for successive values of the relative compression $\delta$. (g) Amplitudes ($A_1$, $A_2$) and wavelength, $\lambda$, as a function of the compression ratio $\delta$. Experimental data for system~1 are reported with symbols {\Large$\bullet$}, {\Large$\filledsquare$} and {\Large$\filleddiamond$} for 30 min., 1h and 2h of irradiation respectively whereas the symbol $\medtriangledown$ is used for system~2. Results of the linear (dotted lines) and nonlinear (solid lines) theories are also reported. Before period-doubling, the expression of the amplitude $A$ computed from Eqs.~(\ref{inext}) is: $\frac{A}{\lambda_0}=\frac{\sqrt{\delta}}{\pi}\left(1-\frac{3\delta}{8}-\frac{17\delta^2}{128} \right)$. The wavelength $\lambda$ is computed from (\ref{link-wavelength}): $\frac{\lambda}{\lambda_0}=1-\delta +O(\delta^3,(B/\lambda_0)^2\delta^2)$, where $B$ is the amplitude of the subharmonic mode.}
\label{fig01}
\end{figure*}

A PDMS film, stretched and then cured with UV/ozone, or a thin polymer film bound to an elastomer foundation, remains initially flat. Under a slight compression, $\delta = (L_0-L)/L_0$, these systems instantaneously forms regular (sinusoidal) wrinkles with a well-defined wavelength, $\lambda_0$. Increasing $\delta$ generates a continuous increase of the amplitude of the wrinkles and a continuous shift to lower wavelength ($\lambda = \lambda_0 (1-\delta)$ see Fig.~\ref{fig01}g). By further compression of the sheet, more complex patterns emerge. Above some threshold, $\delta > \delta_2 \simeq 0.2$, we observe a dramatic change in the morphology leading to a pitchfork bifurcation: one wrinkle grows in amplitude at the expense of its neighbours (Fig.~\ref{fig01}). The profile of the membrane is no longer described by a single cosinusoid but requires a combination of two periodic functions, $\cos \frac{2\pi x}{\lambda}$ and $\cos \frac{2\pi x}{2\lambda}$. The amplitude of the $2 \lambda$ mode increases with the compression ratio, while the $\lambda$ mode vanishes. This effect is similar to period-doubling bifurcations in dynamical systems~\cite{feig78,feig79} observed in, for example, Rayleigh-Bernard convections~\cite{libc82}, dynamics of the heart tissue \cite{guev81,fox02,berg07}, oscillated granular matter~\cite{melo95,venk98} or bouncing droplets on soap film~\cite{gile09}. In contrast to previous works, we describe here a \emph{spatial period-doubling instability} which is rarely observed~\cite{lose96}. Nonlinear coupling between two modes, one with double the wavelength of the other, also appears in post-buckling of cylindrical shells as reported in the classical work of Koiter (see \cite{koit70} and references therein).

The thin inextensible membrane of length $L_0$ is compressed horizontally by a distance $\Delta=L_0-L$ along the $x$-axis and is bound to an elastic foundation that initially fills the half-space $y<0$.  The system is assumed to remain invariant in the $z$ direction (see Fig.~\ref{fig01}). The projected length along the $x$-axis, $L_0-\Delta$, is given by
\begin{equation}
\label{inext}
L_0-\Delta=\int_0^{L_0} d\ell \cos \phi, 
\end{equation}
where $\ell$ is the arc length measured along the curve. The quantity $\phi$ is the angle between the tangent to the surface and the horizontal. The derivative of this angle with respect to the arc length, $\partial_{\ell} \phi$, gives the local curvature of the membrane (for clarity, partial derivatives such as $\frac{\partial}{\partial \ell}$ are written as $\partial_{\ell}$). The relative compression ratio is given by $\delta = \Delta/L_0$.

The response of this thin membrane resting on an elastomer substrate is determined through minimization of the energy per unit of width, $U$. Two energetic contributions are to be considered: \emph{i}) the elastic bending energy of the thin sheet, 
\begin{equation}
\label{bend-energy}
U_B=\frac{B_m}{2}\int_0^{L_0} d\ell (\partial_{\ell} \phi)^2, 
\end{equation}
where the parameter $B_m$ is the bending stiffness of the membrane ($B_m \sim E_m h^3$, $E_m$ being its Young's modulus and $h$ its thickness); 
\emph{ii}) the energy of deformation of the elastomer. The constraint of inextensibility of the membrane (\ref{inext}) is taken into account with the help of a Lagrangian multiplier $F$ identified to the cross-sectional pressure per unit length. The Euler-Lagrange equation obtained from the energy of the system gives the equilibrium of normal forces along the membrane and is given by
\begin{equation}
\label{eqmotion-gen}
	B_m \partial_{\ell}^4 y+F \partial_{\ell}^2 y+P_y=0,
\end{equation}
where $y$ and $P_y$ are functions describing the vertical elevation of the membrane and the normal pressure from the elastomer acting on the membrane, respectively. At linear order, $P_y= K {\cal H}(\partial_{\ell} y)$, where $K$ is the stiffness coefficient of the foundation proportional to its Young's modulus ($K = 2E (1-\sigma)/(1 + \sigma)(3 - 4\sigma)$, where $\sigma$ is the Poisson ratio, see Supplementary Information) and ${\cal H}$ is the Hilbert transform. The first nonlinear contribution due to the elastomer can be computed for periodic deformation with one mode of frequency $q$ and Eq.~(\ref{eqmotion-gen}) reduces then to
\begin{equation}
\label{eqmotion}
	B_m \partial_{\ell}^4 y+F \partial_{\ell}^2 y+K q y + K_2 q^2 y^2=0,
\end{equation}
where $K_2$ is also proportional to the Young's modulus ($K_2 = E (1-2\sigma)(13-16\sigma)/2(1 + \sigma)(3 - 4\sigma)^2$, see Supplementary Information). Notice that, as for the linear response of the susbtrate, the nonlinear term involves also an Hilbert transform for multimode profile. Due to the quadratic nonlinearity from the foundation, the equation \ref{eqmotion} giving the profile of the membrane implies an up-down symmetry breaking: vertical extension and compression along the $y$-axis are no longer equivalent. This equation can also be viewed as a spatial equivalent of a nonlinear oscillator, like a simple pendulum, with which it shares many similarities. 

\begin{figure*}[!hbtp]
\centerline{\includegraphics[width=18.3cm,clip]{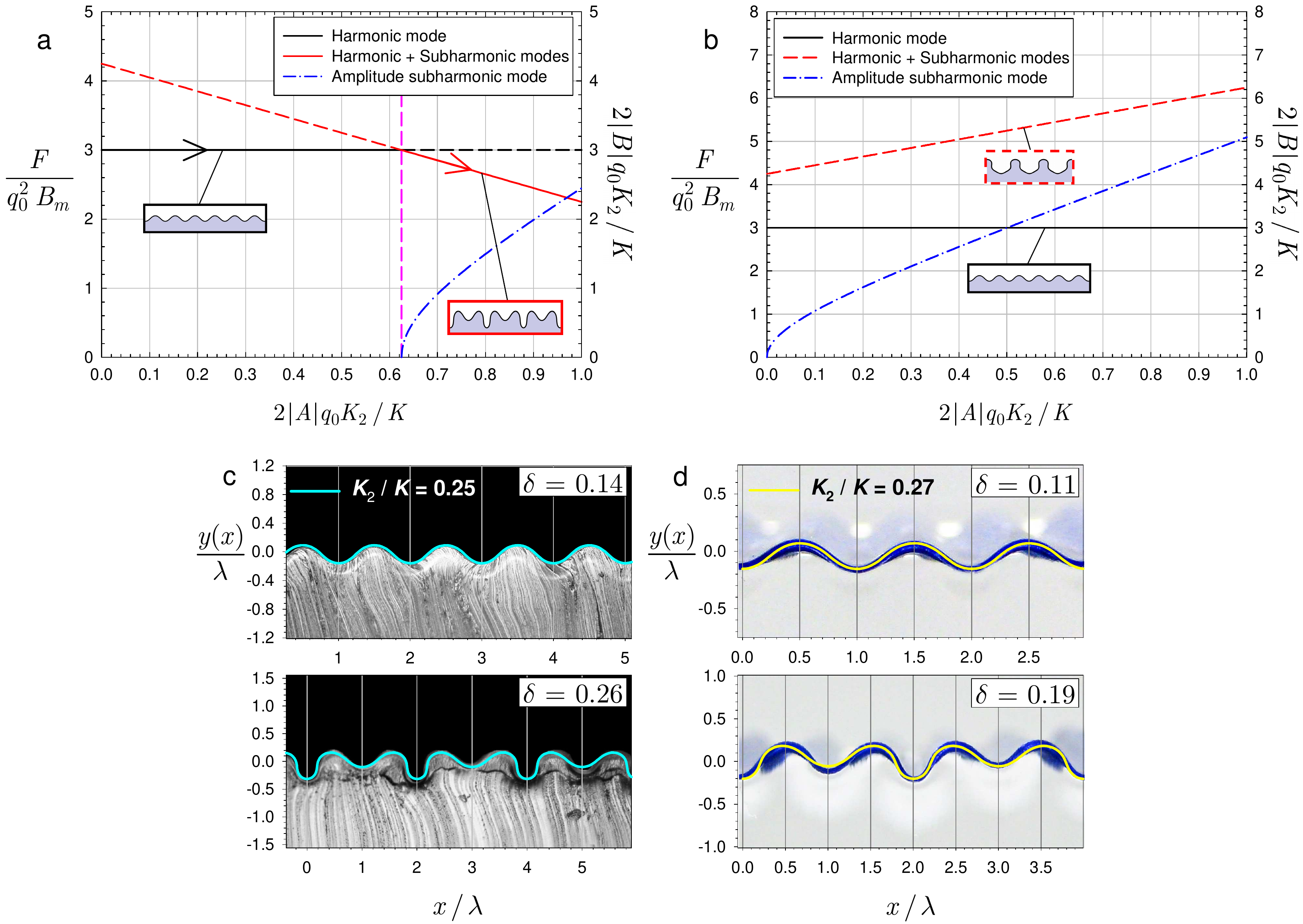}}
\caption{{\bf Predictions of the model and comparison with experimental profiles.} (a), (b) Evolution of the cross-sectional pressure $F$ and of the amplitude $B$ of the subharmonic mode as a function of the amplitude $A$ of the harmonic mode. Arrows indicate the path followed by the systems during compression. (Insets) Representative shapes of the membrane according to the value of $F$. (c) Comparison between theoretical and experimental profiles of the membrane for system 1 for two values of $\delta$. (d) Comparison between theoretical and experimental profiles of the membrane for system 2 for two values of $\delta$.} 
\label{fig02}
\end{figure*}

Equation (\ref{eqmotion}) reduces to a linear oscillator for small amplitudes of the instability. In this regime, the period is independent of the amplitude, as for a simple pendulum, in agreement with observation and usual theories. Indeed, nonlinear terms can be neglected for small amplitudes and the curvilinear and Cartesian coordinates coincide: $\ell\simeq x$, $\phi\simeq \partial_x y$. Equation~(\ref{eqmotion}) admits sinusoidal solutions $y(x)=A \cos(2 \pi x/\lambda)$ provided that the pressure $F$ and the wavelength of the wrinkling instability are related by 
\begin{equation}
	\label{link-pq}
	F(\lambda)=\frac{4\pi^2 B_m}{\lambda^2} +\frac{\lambda K}{2\pi}.
\end{equation}
This relation shows that below a threshold $F<F_c=3q_0^2 B_m$ there is no associated wavelength and the membrane stays flat. At the threshold, $F=F_c$, the wrinkling instability emerges and a unique and constant wavelength, $\lambda_0$, is selected \cite{groe01,cerd03}
\begin{equation}
	\label{wavelength-lin}
	\lambda_0= 2\pi \left(\frac{2B_m}{K}\right)^{\frac{1}{3}} \sim h \left(\frac{E}{K}\right)^{\frac{1}{3}}.
\end{equation}
The selection of this particular wavelength is obtained from a minimization of the energy through a minimization of $F$. The inextensibility constraint (\ref{inext}) gives the evolution of the amplitude of the instability as a function of the relative compression, $A=\pm \lambda_0\sqrt{\delta}/\pi$. However, neither the evolution of the wavelength with $\delta$ nor the period-doubling bifurcation are captured by this linear model.

To determine the supercritical morphology, we study the stability of the single wavelength pattern in the weakly nonlinear regime. We thus consider a small periodic perturbation, $\epsilon u$, characterized by a frequency $k$, of the nonlinear solution for the shape of the membrane: $y\to y + \epsilon u$, $\epsilon$ being arbitrarily small. The equation for the perturbation, $u$, in the leading order in the amplitude $A$ of the instability is then given by
\begin{equation}
	\label{forcing}
	B_m \partial_{\ell}^4 u+F \partial_{\ell}^2 u+K k u = -2 K_2 k A q_0 \cos(q_0 \ell) u.
\end{equation}
The term appearing in the right-hand side of this equation is due to the quadratic nonlinearity of the foundation (stiffness $K_2$). 
Interestingly, this equation is similar to the Mathieu equation, describing resonance in parametric oscillators \cite{mcla62,blan72,sanm84,vand00}. 
For usual forced oscillators, like a simple pendulum with a variable length (the most famous example of this resonance is given by the giant censer, O Botafumeiro \cite{sanm84}), the unforced system is characterized by a given period and the additional frequency needed to produce a parametric resonance is provided by an external agent. 
For all amplitudes of the forcing, the resonance appears provided that forcing and oscillator frequencies are related through $k = q_0/2$.

In our system however, we should also consider a constraint related to the minimization of $F$ ({\it i.e.}, minimization of energy since $U(\delta) = L_0 \int_0^{\delta} F(\delta') d\delta'$ where $\delta$ is the relative compression) determining the amplitude of the forcing term at which the $2\lambda$ mode emerges. Actually, the period-doubling instability cannot be observed for amplitudes smaller than a threshold ({\it i.e.}, defining a compression threshold, $\delta_2$).

\begin{figure*}[!hbtp]
\centerline{\includegraphics[width=18.3cm,clip]{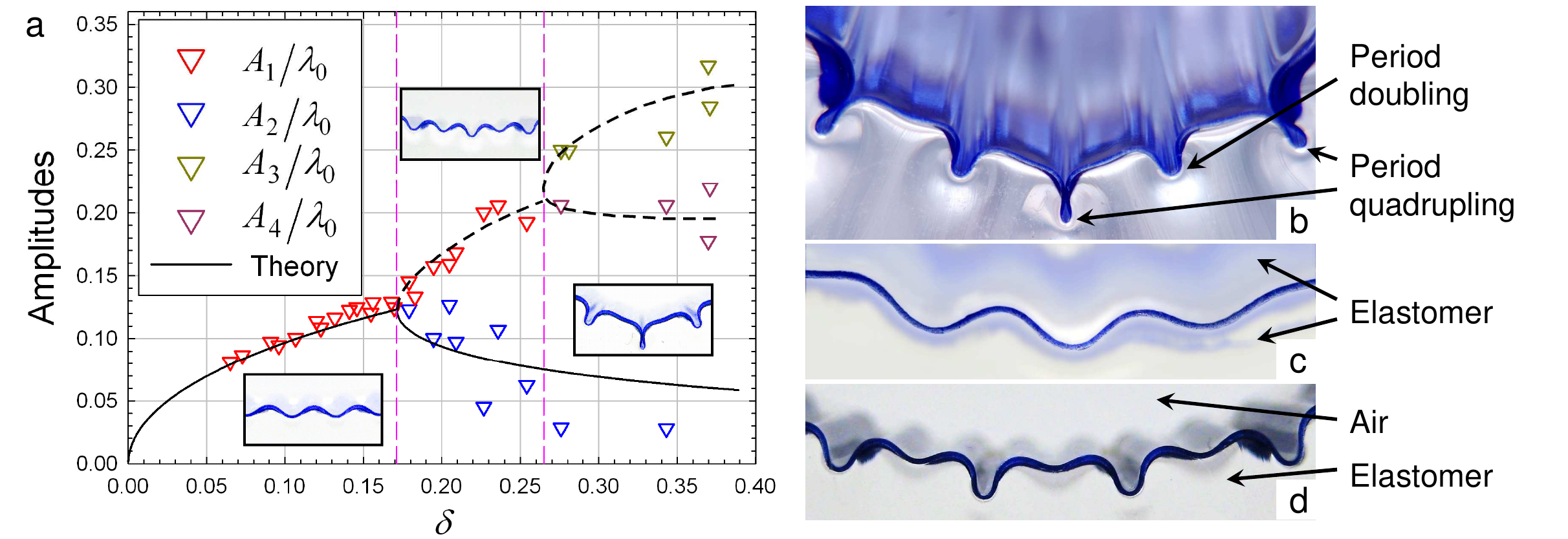}}
\caption{{\bf Additional consequences of the up-down symmetry breaking on wrinkled patterns.} (a) Evolution of amplitudes for large compressions showing the period-quadrupling bifurcation in system 2. The solid curve is obtained numerically for $K_2/K=0.27$. The dashed curve is added to help visualizing the second bifurcation characterizing the period-quadrupling instability. (b) Wrinkled structure showing the period-quadrupling instability ($\delta \simeq 0.37)$. (c) Profile of a thin stiff PDMS membrane resting in between two identical soft PDMS foundations for a relative compression $\delta\simeq 0.23$ (d) Profile of a thin stiff PDMS membrane bound to a soft PDMS foundation for a relative compression $\delta \simeq 0.23$.} 
\label{fig03}
\end{figure*}

From equation (\ref{forcing}), we can deduce that the profile should be described by a multimode solution of the form, $y(\ell) = \sum_{k=1}^{\infty} C_k \cos (k q_0\ell/2)$. Indeed, without any loss of generality, the wrinkled pattern can be assumed to be described by an even function since the system is invariant under horizontal translation. 
The numerical analysis of equation (\ref{eqmotion}), adapted to multimode periodic solutions, shows a very good agreement with experimental data (Fig.~\ref{fig01}g). Notice that the convergence is already reached with the four first modes (see Supplementary Information). The relevance of the model is further demonstrated by the excellent agreement between experimental and calculated profiles (Figs.~\ref{fig02}c and d). We should emphasize that the model relies on a single parameter, $K_2/K$, that determines the period-doubling threshold $\delta_2$.


In order to preserve an explicit analysis and to capture the physics of the model, we restrict the following discussion to the ansatz $y(\ell)=A\cos(q_0\ell) + B \cos(q_0 \ell/2)$. Substituting this ansatz in Eq.~(\ref{eqmotion}), we obtain a system of two equations in $A$, $B$ and $F$, admitting two solutions. A trivial solution corresponds to the evolution before period-doubling: $F/(q_0^2 B_m)=3$ and $B=0$ ($A$ being determined by the inextensibility constraint). The second solution involving a subharmonic mode ($B \neq 0$) reads
\begin{eqnarray}
	\label{eq-f}
	\bar{F}&=&17/4 +  2 \bar{A}, \\
	\label{amp-b}
	\bar{B}^2&=& 2 \bar{A}\left(5 + 8 \bar{A}\right).
\end{eqnarray}
where $\bar{F} = F/(q_0^2 B_m)$, $\bar{A}=2K_2 A q_0/K$ and $\bar{B}=2K_2 B q_0/K$. Equation (\ref{eq-f}) is no longer invariant under a change of sign of $A$. Indeed, the amplitude $A$ of the harmonic mode can be either positive or negative since the nonlinear system is characterized by a up-down symmetry breaking due to the quadratic nonlinearity of the foundation upon deformation. 
Fig.~\ref{fig02}a and b shows the evolutions of both solutions with the amplitude, $\bar{A}$.
The symmetry breaking induces two regimes. For $\bar{A}>0$, $\bar{F}$ is always larger than the value associated to the harmonic mode alone, {\it i.e.} $\bar{F}=3$. The corresponding shape for the membrane is forbidden and thus not observed experimentally, see Fig.~\ref{fig02}b. In contrast, for $\bar{A}<0$, the emergence of a subharmonic mode is energetically favorable ($\bar{F}<3$) beyond a threshold value, see Fig.~\ref{fig02}a. From Eq.~(\ref{amp-b}), we observe that $\bar{B}$ starts to grow precisely from this threshold. This analysis does not, however, imply that an harmonic mode with a positive amplitude, $\bar{A}>0$, is stable against subharmonic perturbations. Indeed, the above analysis is performed using, without loss of generality, an even function to describe the evolution of the wrinkled pattern. Having found the energetically favorable pattern in this case, we can use the translation invariance to generate equivalent patterns: $y(\ell-\pi/q_0)=-A\cos(q_0\ell) + B \sin(q_0 \ell/2)$. The sign of $A$ being now reversed, it implies that an harmonic mode with a positive amplitude is also unstable against subharmonic perturbations above the same thershold and leads to the same wrinkled pattern but translated.

Through the inextensibility constraint, the threshold for $\bar{A}$ implies the existence of a critical relative compression, $\delta_2$, for the onset of the period-doubling instability. Using the relation between $A$ and $\delta$ at the lowest order, we obtain
\begin{equation}
	\label{delta2}
	\delta_2=\left(\frac{5}{32(K_2/K)}\right)^2 \simeq 0.02 \frac{(1-\sigma)^2}{(1-2\sigma)^2}.
\end{equation}
The critical compression needed to observe a period-doubling bifurcation for wrinkling instability, $\delta_2$, strongly decreases with the Poisson ratio of the elastic foundation. The values found numerically for the ratio $K_2/K \sim 0.25$ yield a Poisson ratio around 0.44 which is close to the value usually reported in literature for PDMS ($\sim 0.48$).

Moreover, this model based on nonlinear oscillator should imply that, for larger amplitudes of the $2 \lambda$ mode, a period-quadrupling bifurcation characterized by a wavelength $4 \lambda$ would appear. This behaviour is indeed observed in Fig.~\ref{fig03}a, b for compression ratios larger than $0.26$. This last observation clearly suggests that cascades of \emph{spatial period-doubling bifurcations} can be observed for the elastic instability of rigid membrane, provided that the up-down symmetry is broken. Such a cascade is known to lead to chaos after several bifurcations \cite{feig78,feig79}. There is however a geometric limitation in our system in contrast to previously reported temporal period-doubling cascade. Indeed, the evolution of the pattern saturates as soon as sharp folds appear (see Fig.~\ref{fig03}b). For instance, due to finite thickness of the membrane, we experimentally reached at most period-quadrupling structures.

A further confirmation of our approach can be obtained. Our interpretation of the period-doubling bifurcation in rigid membrane on elastomer implies that the dynamics should be governed by nonlinear terms of even order, which break the up-down symmetry. Consequently, systems with an up-down symmetry, like a thin elastic membrane resting on a liquid~\cite{poci08}, do not develop a period-doubling instability. Interestingly, we could make trilayers restoring the symmetry. Indeed, a system composed of a thin elastic membrane in-between two identical soft foundations, one below and one above the membrane, does not exhibit the period-doubling bifurcation. Instead it develops patterns similar to those observed with floating membranes. In Fig.~\ref{fig03}b and c, we compare the profile of the membrane when there are one or two foundations for the same compression ratio. 

The second salient feature of the nonlinear wrinkling instability is the continuous decrease of the wavelength with the compression ratio $\delta$. This effect arises from the change from curvilinear to Cartesian coordinate. The wavelength is measured along the horizontal $x$-axis while the shape of the membrane is determined in curvilinear coordinates $\ell$ where it is constant. For a periodic profile $y(\ell)$, with a wavelength $\lambda_{\ell}$, $\lambda_x\equiv \lambda$ is given by
\begin{equation}
	\label{link-wavelength}
	\lambda= \int_{0}^{\lambda_{\ell}} d\ell \cos \phi = \int_{0}^{\lambda_{\ell}} d\ell \, \sqrt{1-(\partial_{\ell} y)^2}.
\end{equation}
The evolution of the wavelength along the horizontal $x$-axis at the leading order in the amplitude of the instability, $A$, is given by 
\begin{equation}
	\label{evol-wavelength-d}
	\frac{\lambda}{\lambda_0} = 1-\frac{(A \pi)^2}{\lambda_0^2}= 1-\delta
\end{equation}
in very good agreement with experimental data in Fig.~\ref{fig01}g.

The universal model describing the formation of wrinkled patterns based on nonlinear oscillator dynamics should explain observations in very different fields. 
For example, a better understanding of the elastic instability of rigid membranes will help to determine the exact mechanisms leading to the growth of wrinkled morphology in living systems. 
It is also a new blueprint to develop multiple-length-scale microfabrication techniques useful in the design of specific topography.\\

{\bf Methods}

Experiments were carried out using polydimethylsiloxane (PDMS) elastomer (Sylgard 184) purchased from Dow Corning. Two different systems were studied. System 1: a bare elastomer of PDMS is irradiated with UV in presence of oxygen. Ozone is generated and will affect the crosslinks density of the PDMS outer surface. The rigidity of the surface drastically increases with the irradiation time to finally yield a brittle overlayer covalently bound to the uncured elastomer. System 2: multilayers prepared by a simple assembly of monolayers of different elastic properties. The ``rigid'' and ``soft'' layers correspond to elastic modulus values of 1200 and 10 kPa, respectively. To ensure a very strong adhesion between both PDMS films and avoids delamination during the compression, these two PDMS elastomers were assembled by contact after a plasma curing (in a Plasma Cleaner oven). The experimental set-up was a custom-built stretching/compressing device. The UV/O3 modified PDMS was compressed by using a stretched/curing/release experiments. The measurements were achieved using image analysis from microtomed slices of the samples. The bilayer PDMS assembly were compressed by inducing a macroscopic radius of curvature. The measurements were performed from macro photography of the cross-section of the samples (see Supplementary Information for further details).

{\bf Acknowledgements}  The authors thank T. Witten, B. Davidovitch, H. Diamant, S.~Desprez, C.~Troetsler, S.~Gabriele and G.~Carbone for fruitful discussions. 
This work was supported by the Belgian National Funds for Scientific Research (Mandat Impulsion Scientifique), the Government of the Region of Wallonia (CORRONET and REMANOS Research Programmes) and the European Science Foundation (Eurocores FANAS programme, EBIOADI collaborative research project). 
F.B. acknowledges financial support from a return grant delivered by the Federal Scientific Politics.

\clearpage

{\bf {\it Supplementary information for} ``Multiple-length-scale elastic instability from period-doubling bifurcation cascade mimics parametric resonance of nonlinear oscillators''}

\section{Materials and methods}

\begin{figure}
\centerline{\includegraphics[width=6cm,clip]{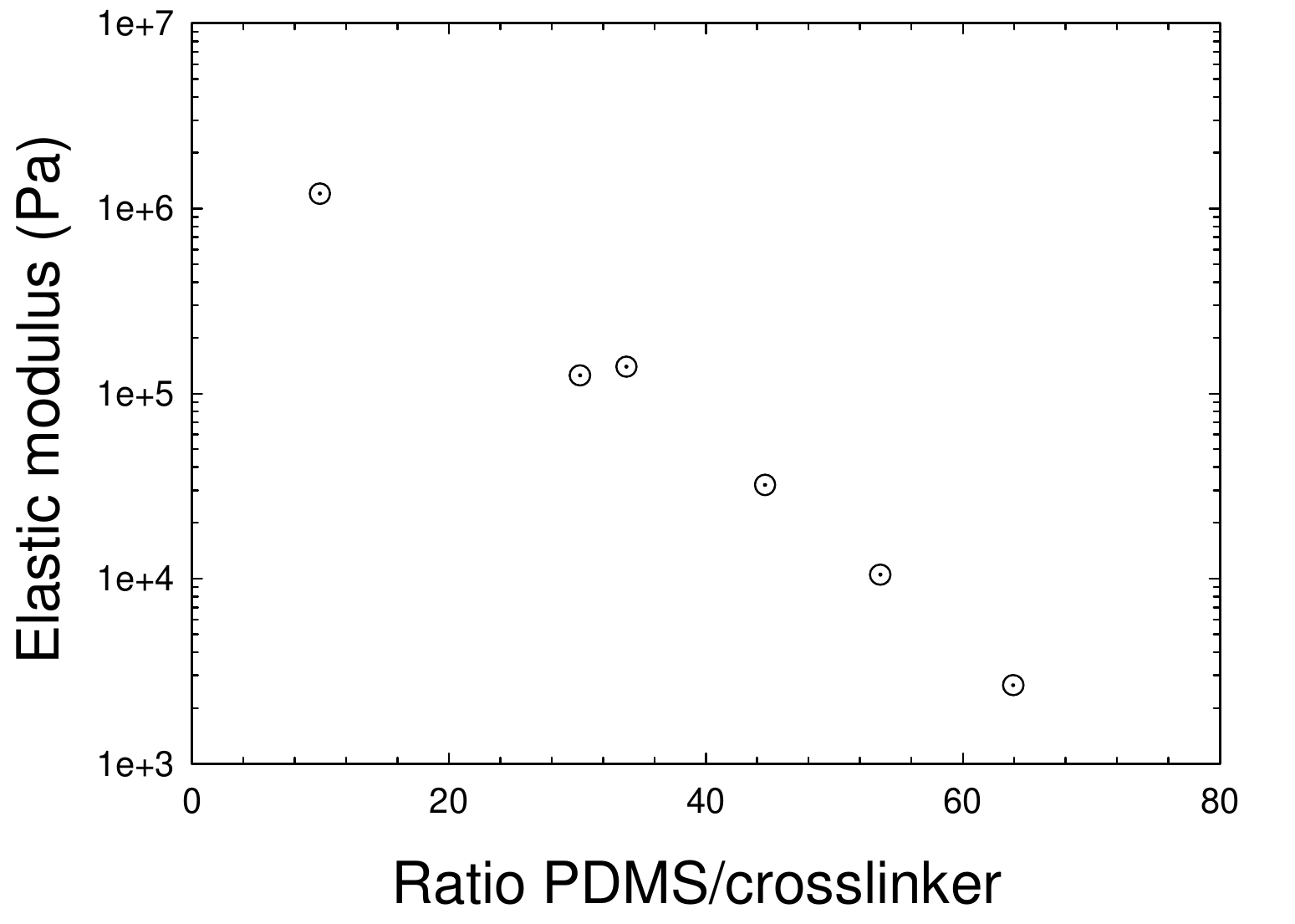}}
\caption{Evolution of the elastic modulus of the PDMS elastomers as a function of the fraction of crosslinker (the fraction recommended by Dow Corning is 10\% to obtain the more rigid elastomers).} 
\label{crosslink}
\end{figure}

Experiments were carried out using polydimethylsiloxane (PDMS) elastomer (Sylgard 184) purchased from Dow Corning. By changing the proportion of crosslinker, we have adjusted the elastic properties of the elastomer. The elastic moduli measured for various crosslinked PDMS elastomers is given in Fig.~\ref{crosslink} [1].

Two different systems were studied: \emph{i}) UV/O3 cured PDMS films (system 1) and, \emph{ii}) multilayers prepared by a simple assembly of monolayers of different elastic properties (system 2). 

By irradiating a bare elastomer of PDMS with UV in presence of oxygen, we generate ozone molecules that modify the crosslinks density of the PDMS outer surface. As demonstrated by numerous studies~[2], the rigidity drastically increases with the irradiation time to finally yield a brittle overlayer covalently bound to the uncured elastomer. As showed below, the thickness of modified PDMS layer increases with irradiation time. 

In contrast, the multilayers systems were prepared by assembling two PDMS films of different rigidity (high/low crosslinks density with high/low elastic modulus). The ``rigid'' and ``soft'' layers correspond to elastic modulus values of 1200 and 10 kPa, respectively. 
After a plasma curing during four minutes (in a Plasma Cleaner oven), these two PDMS elastomers were assembled by contact. The plasma curing ensures a very strong adhesion between both PDMS films and avoids delamination during the compression of the multilayer.

The experimental set-up was a home-made stretching/compressing device. 
The UV/O3 modified PDMS was compressed by using a stretched/curing/release experiments (maximum compression ratio, $\delta \simeq 0.55$). The measurements were achieved from image analysis from microtomed slices of the samples perpendicular to the symmetry axis of the pattern. Due to the very small wavelength observed for UV/O3 systems ($10 - 100\, \mu$m) a microscope was required. The bilayer PDMS assembly were compressed by inducing a macroscopic radius of curvature (maximum compression ratio, $\delta \simeq 0.4$). The measurements were performed from macro photography of the cross-section of the samples.

\begin{figure}
\centerline{\includegraphics[width=6cm,clip]{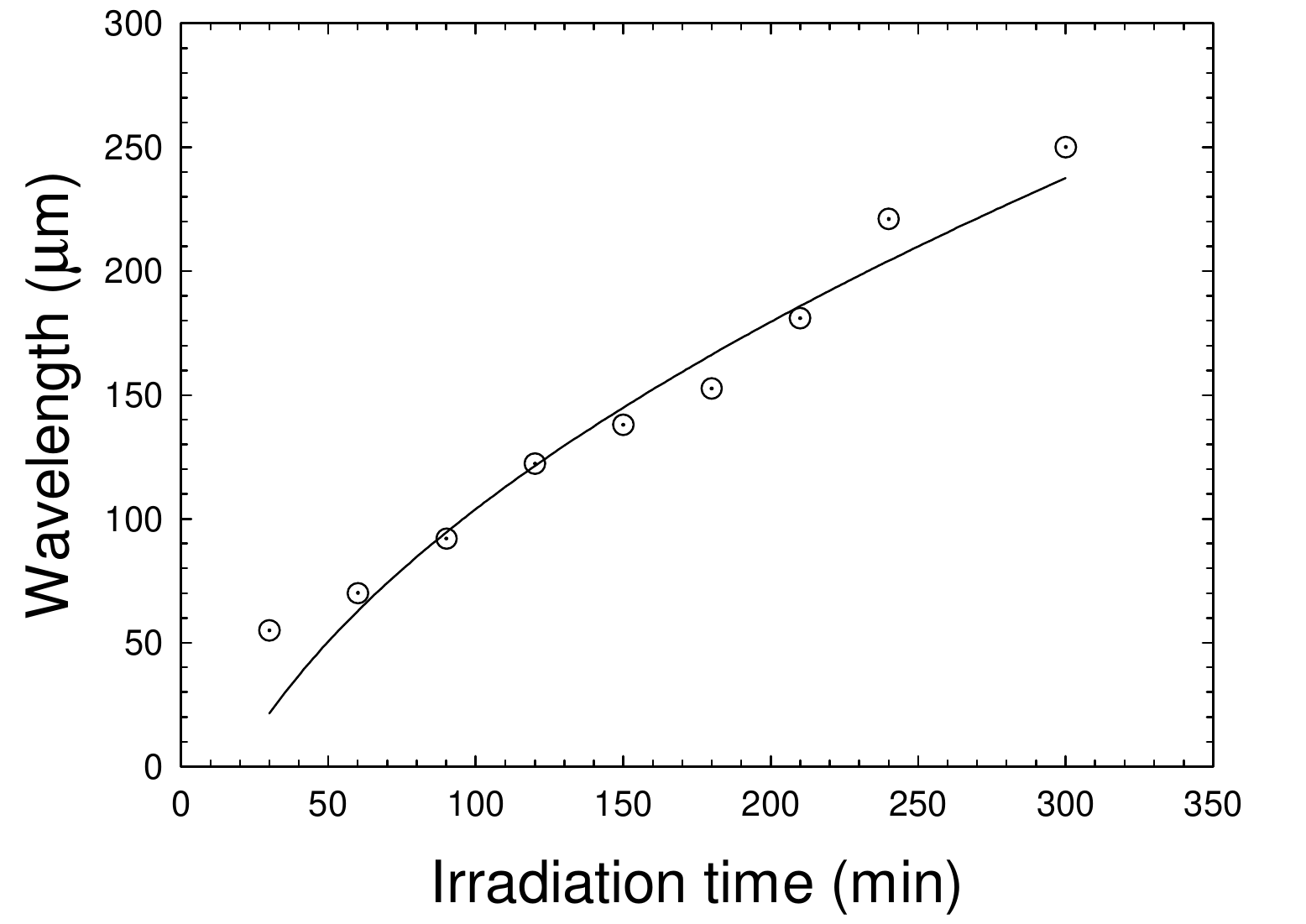}}
\caption{Evolution of the wavelength at the threshold, $\lambda_0$, with the curing time in UV/O3 oven (irradiation time). The solid line corresponds to a fit of the data with a rational scaling law ($\lambda_0 \propto \sqrt{t}$).} 
\label{lambda}
\end{figure}

\section{Influence of UV/O3 on PDMS elastomers}

As demonstrated by several studies, UV/O3 curing of PDMS generates a rigid overlayer by a chemical transformation of the PDMS chains, the chemical composition for very long irradiation time corresponding to brittle silica~[2]. 
In addition, the wrinkling instability can be used to check the finite size of the modified PDMS surface (Fig.~\ref{lambda}).

Considering the relation  $\lambda_0 \propto h (E_m/E)^{1/3}$ (where $h$, $E_m$ and $E$ are the thickness, the elastic modulus of the rigid ``membrane'' and the elastic modulus of the foundation, respectively), the evolution of the initial wavelength with irradiation time is in agreement with a diffusive behaviour. Indeed, the UV/O3 curing affects the outer surface of the PDMS film up to an effective thickness, $h_{\text{eff}}$, expected to grow with the irradiation time according a $\sqrt{t}$ law. 

\section{Parametrization, equation for the membrane and origin of the nonlinearity}

The membrane is assumed to be inextensible and invariant along one direction such that it is completely determined by its $(x,y)$ Cartesian coordinates. Let $\ell$ be the curvilinear coordinate and $\phi$ the angle between the tangent to the surface and the horizontal direction. 
According to elementary differential geometry, the shape of the surface is then parametrized by
\begin{eqnarray}
\label{param}
(x,y)&=&\left(\int_0^{\ell} ds \cos \phi, \int_0^{\ell} ds \sin \phi \right), \nonumber \\
     &=&\left(\int_0^{\ell} ds \sqrt{1-(d_s f(s))^2}, f(\ell) \right),
\end{eqnarray}
where $f$ gives the height of the membrane as a function of the curvilinear coordinate.
In addition, the derivative $\partial_{\ell} \phi$ is equal to the curvature of the membrane.

The thin rigid membrane resting on the soft thick foundation is assumed to be invariant along one direction. Consequently, the evolution of the system under compression is described in a plane $(x,y)$. For small deformations, the equation giving the shape of a free and inextensible plate resting initially at $y=0$ and invariant along one direction reads~[3]
\begin{equation}
	\label{tige}
	B_m y^{(IV)} + F y'' =0,
\end{equation}
where $B_m = E_m h^3/(12(1-\sigma^2))$ is the flexural rigidity. $E_m$ is the Young modulus, $\sigma$ is the Poisson coefficient and $h$ is the thickness of the plate. $F$ is the in-plane applied load on the plate. For large deformations, the nonlinear terms are of odd order and the first nonlinearity is cubic. Indeed, this system has an up-down symmetry: bending up or down a plate is energetically equivalent. 

If an elastomer fills the semi-plane $y<0$, an additional term, describing the force normal to the surface due to the elastomer, must be added to Eq.~(\ref{tige}). For small deformations, the linear theory of elasticity can be used. The equation satisfied by the displacement vector, $\vec{u}$, is given by~[3] 
\begin{equation}
	\label{u-lin}
	(1 - 2\sigma )\Delta \vec{u}  + \vec{\nabla} \left( {\vec \nabla  \cdot \vec u} \right) = 0,
\end{equation}
where $\sigma$ is the Poisson ratio. We assume that there is no deformation along the $z$-axis; the deformation is then studied in the $(x,y)$-plane. Under this condition, Eq.~(\ref{u-lin}) reduces to
\begin{eqnarray}
	\label{system}
	2(1-\sigma)\partial_x^2 u_x+(1-2\sigma)\partial_y^2 u_x + \partial_{x y}^2 u_y &=& 0, \nonumber \\
	2(1-\sigma)\partial_y^2 u_y+(1-2\sigma)\partial_x^2 u_y + \partial_{x y}^2 u_x &=& 0.	
\end{eqnarray}
The boundary conditions are $u_i(x,y\to -\infty) = 0$ ($i=x,y$), $u_x(x,y=0)=0$ and $u_y(x,y=0)=f(x)$, where $f(x)$ is a known function describing the shape of the surface of the foundation. The vertical force per unit area at the surface of the elastic substrate subject to deformation $f(x)$ is given by~[3]
\begin{equation}
	\label{force}
	P_y(x,y=0)=\sigma_{y i}\, n_i,
\end{equation}
where there is summation on repeated indices and where $\hat{n}$ is the unit normal vector to the surface. At the lowest order in the amplitude of $f(x)$, we simply have $\hat{n}=(0,1)$ and $P_y$ is given by $\sigma_{yy}(x,y=0)$. To compute this quantity, we solve Eq.~(\ref{system}) with the appropriate boundary conditions given above. Once the displacement vector is known, the strain tensor is computed, in linear theory of elasticity, from~[3]
\begin{equation}
	\label{strain-lin}
	u_{ik}=\frac{1}{2}\left(\frac{\partial u_i}{\partial x_k} + \frac{\partial u_k}{\partial x_i}\right).
\end{equation}
The stress tensor is then given by
\begin{equation}
	\label{stress}
	\sigma_{ik}=\frac{E}{1+\sigma}\left(u_{ik}+\frac{\sigma}{1-2\sigma}u_{ll} \delta_{ik} \right).
\end{equation}
To solve Eq.~(\ref{system}), we use the Fourier transform of $u_i(x,y)$ along the $x$-axis
\begin{equation}
	\tilde{u}_i(k,y)={\cal F}(u_i(x,y))=\frac{1}{\sqrt{2\pi}}\int_{-\infty}^{\infty} u_i(x,y) e^{ikx} dx.
\end{equation}
The PDE system (\ref{system}) reduces then to this ODE system
\begin{eqnarray}
	(1-2\sigma)d_y^2 \tilde{u}_x -ik d_y \tilde{u}_y-2(1-\sigma)k^2 \tilde{u}_x&=&0, \nonumber \\
	2(1-\sigma)d_y^2 \tilde{u}_y-ik d_y \tilde{u}_x- (1-2\sigma)k^2 \tilde{u}_y&=&0.
\end{eqnarray}
This last system of equations can be decoupled to obtain
\begin{eqnarray}
	d_y^4 \tilde{u}_x - 2 k^2 d_y^2 \tilde{u}_x + k^4 \tilde{u}_x&=&0, \nonumber \\
	ik d_y \tilde{u}_y +2(1-\sigma)k^2 \tilde{u}_x -(1-2\sigma) d_y^2 \tilde{u}_x &=&0.
\end{eqnarray}
The solution which satisfies the boundary conditions reads
\begin{eqnarray}
	\label{sol-fourier}
	\tilde{u}_x(k,y)&=&\frac{ik \tilde{f}(k)}{(3-4\sigma)}\, y\, e^{|k|y}, \nonumber \\
	\tilde{u}_y(k,y)&=&\frac{\tilde{f}(k)}{(4\sigma-3)}(4\sigma-3+|k|y)\, e^{|k|y}.
\end{eqnarray}
Using Eqs.~(\ref{strain-lin}) and (\ref{stress}) in Fourier space, we obtain
\begin{equation}
	\label{p-fourier}
	\tilde{P}_y(k,y=0)=K\, |k|\, \tilde{f}(k),
\end{equation}
where
\begin{equation}
	K=\frac{2(1-\sigma)E}{(1+\sigma)(3-4\sigma)},
\end{equation}
$E$ being the Young's modulus. Consequently, the quantity to be added to Eq.~(\ref{tige}) is given by
\begin{equation}
	\label{p}
	P_y(x,y=0)=K {\cal F}^{-1}(|k|\, \tilde{f}(k)).
\end{equation}
This expression still involves the Fourier transform of $f(x)$. It is possible to obtain an equivalent form involving only $f(x)$. The Hilbert transform, ${\cal H}$, of a function $f(x)$ is a linear operator given, by definition, by the convolution of $f(x)$ and $h(x)=1/(\pi x)$: ${\cal H}(f(x))=(h\star f)(x)$. Using the properties of the Fourier transform we then have
\begin{eqnarray}
	{\cal F}\left({\cal H}(\partial_x f(x)) \right)&=&\sqrt{2\pi} {\cal F}(h) {\cal F}(\partial_x f(x)),\nonumber \\
	&=& \sqrt{2\pi} \left(-\frac{i}{\sqrt{2\pi}}\ \text{sgn}(k) \right)(i k \tilde{f}(k)), \nonumber \\
	&=& \text{sgn}(k) k \tilde{f}(k) = |k| \tilde{f}(k).
\end{eqnarray}
Using this last relation, we can now write
\begin{equation}
	\label{p-final}
	P_y(x,y=0)=K {\cal H}(\partial_x f(x)).
\end{equation}
Due to the property of the Hilbert transform (like ${\cal H}(\cos(qx))= \sin(qx)$ and ${\cal H}(\sin(qx))= -\cos(qx)$), for a periodic profile characterized by only one frequency $q$, this additional term is equivalent to $q y$. Consequently,in the linear regime, the equation to solve for a periodic deformation characterized by only one frequency $q$ is then given by
\begin{equation}
	\label{tige-found}
	B_m y^{(IV)} + F y'' +K q y=0,
\end{equation}
where $K=K(E,\sigma)$ is the stiffness of the elastic foundation.

Now, we extend this result for the response of the substrate to the case of larger deformations where nonlinearities are not negligible. Here, we compute the first nonlinear correction to the response of the foundation restricting our analysis to deformations relevant to our system. The purpose of this paper is to understand the emergence of the subharmonic mode. We thus perform a weakly nonlinear analysis valid up to the threshold of the period-doubling instability.

For larger deformations, the contribution of the elastic foundation must be computed using nonlinear elasticity theory. The relation between the strain tensor $u_{ik}$ and the displacement vector $\vec{u}$ is now given by~[3]
\begin{equation}
	\label{strain-dis}
	u_{ik}=\frac{1}{2}\left(\frac{\partial u_i}{\partial x_k} + \frac{\partial u_k}{\partial x_i}+\frac{\partial u_l}{\partial x_i}\frac{\partial u_l}{\partial x_k}\right),
\end{equation}
where there is summation on repeated index. In linear theory of elasticity, the quadratic term on the right-hand side of (\ref{strain-dis}) is neglected (see Eq.~(\ref{strain-lin})). It is however possible to rederive the equation of elasticity taking into account this nonlinear term since the equation satisfied by the displacement vector is obtained at equilibrium by $\partial \sigma_{ik}/\partial x_k=0$, where the expression of the stress tensor is given by (\ref{stress}). This nonlinear equation reads
\begin{eqnarray}
	\label{u-nonl}
	&&(1 - 2\sigma )\Delta u_i  + \nabla _i \left( {\vec \nabla  \cdot \vec u} \right)  \\ &+& (1 - 2\sigma )\sum\limits_{k\ell } {\frac{{\partial ^2 u_\ell  }}{{\partial x_k^2 }}\frac{{\partial u_\ell  }}{{\partial x_i }}}  + \sum\limits_{k\ell } {\frac{{\partial ^2 u_\ell  }}{{\partial x_k \partial x_i }}\frac{{\partial u_\ell  }}{{\partial x_k }}}  = 0, \nonumber
\end{eqnarray}
with $i=1,2,3$. For a given shape of the deformation of the elastic foundation, the normal force per unit area, $\vec{P}$, is still computed from the relation $P_i = \sigma_{ik} n_k$, where $\vec{n}$ is the unit normal vector to the surface. Once $\vec{P}$ is known, it can be added to (\ref{tige-found}). Setting $\eta=\sqrt{\delta}$, all quantities are expanded in power of $\eta$. Equation (\ref{u-nonl}) is then solved in perturbation with appropriate boundary condition and in particular $u_y(x,y=0)= A \cos(q x) + C \cos(2q x)$, where $A$ is assumed to be of order $\eta$ and $C$ of order $\eta^2$. The nonlinear terms of Eq.~(\ref{u-nonl}) being quadratic, the first nonlinear mode of the wrinkled pattern of our system will be of the form $\cos(2qx)$. This nonlinear shape used as boundary condition should describe accurately enough the profile of our system up to the emergence of the subharmonic mode. Sufficiently near the threshold of the period-doubling instability, the amplitude of this subharmonic mode remains very small with respect to the dominant mode. The nonlinearity associated to this subharmonic mode should thus remain negligible.

The lowest nonlinear response of the elastic deformation to a periodic deformation of the system with frequency $q$ is then found to be $P_y= Kq (A \cos(q x) + 2 C \cos(2q x)) +$ $(K_2/2) A^2q^2 \cos(2qx)$, with $K_2 = E (1-2\sigma)(13-16\sigma)/2(1 + \sigma)(3 - 4\sigma)^2$. At order $\eta^2$, this expression can be written as $P_y=K q y +K_2 q^2 (y^2 - \langle y^2 \rangle)$, where $\langle \cdot \rangle = \lambda^{-1}\int_0^{\lambda} \cdot\, dx$. For our purpose, since we are considering periodic solutions, it can also be written as $P_y=K {\cal H}(y') +K_2 [{\cal H}(y')^2-\langle {\cal H}(y')^2 \rangle$]. To simplify the notations, the average term is dropped in the following equations, and in the main text, but is taken into account in the computations. The resulting nonlinear equation, valid only for wrinkled pattern characterized by only a single mode of frequency $q$, is then found to be
\begin{equation}
	B_m y^{(IV)} + F y'' + K q y + K_2 q^2 y^2=0.
\end{equation}
In this equation, cubic or higher order nonlinear terms, coming from the bending of the membrane or the deformation of the foundation, are neglected. Notice that for multimode profile, characterized by several frequencies $q_i$, the nonlinear equation involving the Hilbert operator should be used.

\section{Period-doubling bifurcation}
\label{app2}

\begin{figure}[!hbtp]
\centerline{\includegraphics[width=6cm,clip]{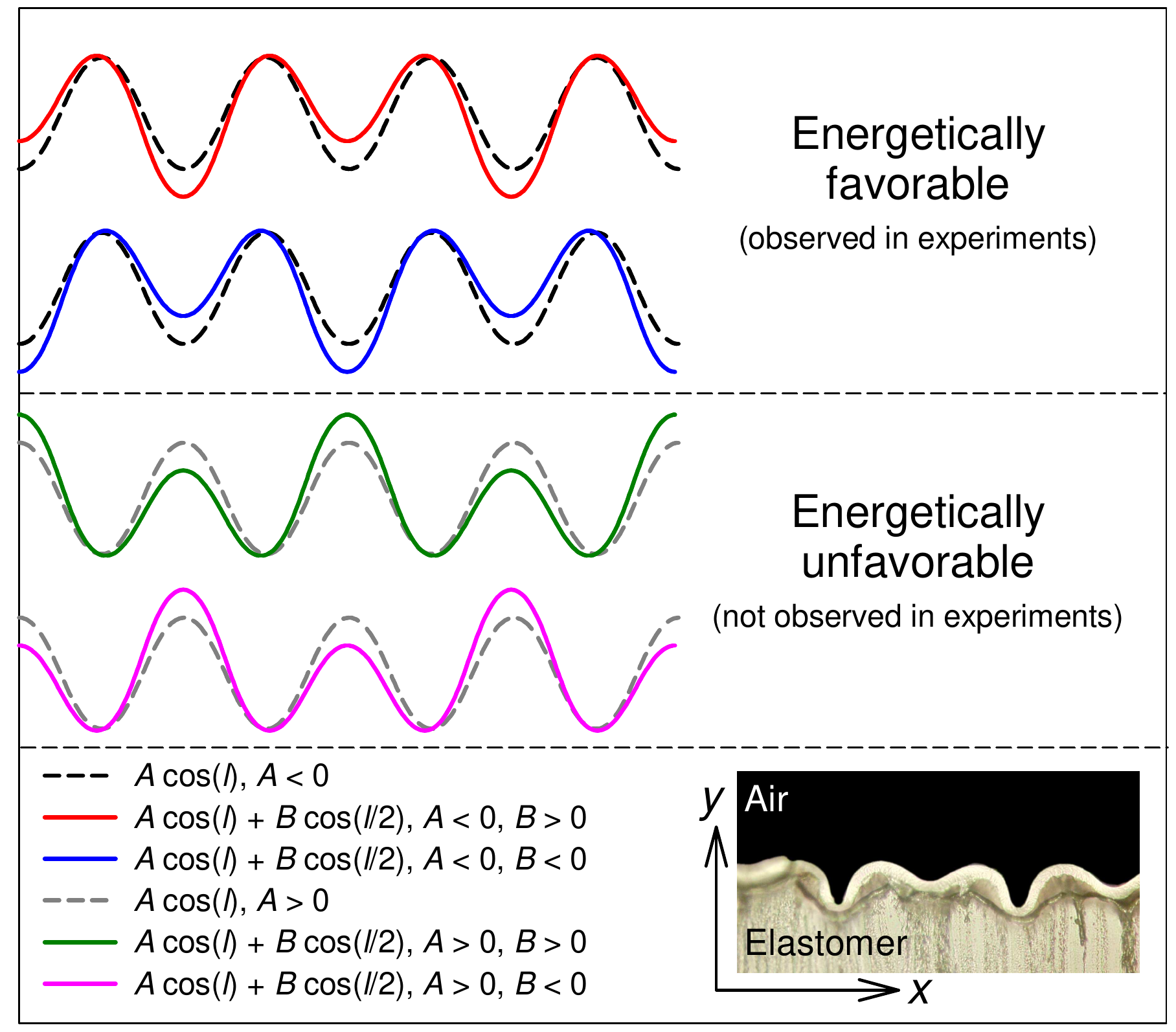}}
\caption{Schematic views of the possible shapes of the membrane after period-doubling according to the sign of the harmonic mode.} 
\label{fig03-sup}
\end{figure}

We consider the equation that the shape of the surface of the system, $y$, must satisfy at the quadratic order
\begin{equation}
	\label{eq-quadra}
	B_m y^{(IV)} + F y'' + K {\cal H}(y') + K_2 {\cal H}(y')^2=0.
\end{equation}
The solution at quadratic order that minimizes the energy, {\it i.e.} minimizing $F$, can be written as
\begin{equation}
	y   = A \cos(q_0 \ell) + \text{super-harmonic terms of order}\, \eta^2,
\end{equation}
with $A=\pm 2 \eta/q_0$, given by the inextensibility constraint and $\eta = \sqrt{\delta}$. We perform now a linear stability analysis of this solution against periodic perturbations, $u$, characterized by a single frequency $k$. We substitute the function $y + \epsilon u$ into Eq.~(\ref{eq-quadra}), $\epsilon$ being an infinitesimal quantity. At the first order in $\epsilon$ and $\eta$ we obtain
\begin{eqnarray}
	\label{forcing-app2-temp}
	B_m u^{(IV)} + F u'' + K {\cal H}(u') &=& - 2 K_2 {\cal H}(y') {\cal H}(u'),  \\
	                                      &=& - 2 K_2 A q_0 \cos(q_0\ell) {\cal H}(u'). \nonumber
\end{eqnarray}
Since $u$ is characterized by a single frequency $k$, this equation can be written as
\begin{equation}
	\label{forcing-app}
	B_m u^{(IV)} + F u'' + K k u = - 2 K_2 A q_0 k \cos(q_0\ell) u.
\end{equation}
This equation is identical to Eq.~(7) reported in the main text. We rescale now the parameters, the independent variable and the function as follow: $F=\bar{F} q_0^2 B_m$, $K=2 q_0^3 B_m$, $K_2=\bar{K}_2 q_0^3 B_m$, $q_0\ell = z$ and $u = \bar{u}/q_0$. Equation~(\ref{forcing-app}) takes now the form
\begin{equation}
	\label{forcing-app2}
	\bar{u}^{(IV)} + \bar{F} \bar{u}'' + 2 \omega \bar{u} = - 2 A q_0 \bar{K}_2  \omega \cos(z) \bar{u},
\end{equation}
where $\omega=k/q_0$. Assuming $\bar{u}=\cos(\omega z)$, we obtain
\begin{eqnarray}
	\left(\omega^3-\bar{F}\omega +2\right)\cos(\omega z)&=& -A q_0\bar{K}_2\left[\cos((1-\omega)z) \right. \nonumber \\ &+& \left. \cos((1+\omega)z)\right].
\end{eqnarray}
If $\omega < 1/2$, keeping the lowest order Fourier modes, the equation is satisfied provided $\bar{F}=\omega^2+2/\omega$. However for these values of $\omega$, $\bar{F}$ is always larger than the value obtained for the harmonic mode, {\it i.e.} $\bar{F}=3$. Consequently, this mode, $\cos(\omega z)$, cannot emerge because the shape adopted by the system is the one that minimizes $F$. If $\omega > 1/2$, the equation cannot be satisfied by keeping only the lowest order Fourier mode. However, if $\omega = 1/2$, the equation is satisfied if 
\begin{equation}
	\label{eq-f}
	\bar{F}=17/4+2A q_0 \bar{K}_2.
\end{equation}
The amplitude of the harmonic mode, $A$, is either positive or negative since the system, at the linear order, has an up-down symmetry. However, adding a quadratic nonlinearity breaks this symmetry, this is why Eq.~(\ref{eq-f}) is no longer invariant under a change of sign of $A$. For large enough (in absolute value) negative values of the amplitude $A$, {\it i.e.} $-A q_0\bar{K}_2 > 5/8$, Eq.~(\ref{eq-f}) shows that adding a subharmonic mode to the shape of the membrane leads to a smaller value of $\bar{F}$ than the value obtained with an harmonic mode alone. This critical amplitude implies a threshold, $\delta_2$, for the period-doubling instability through the inextensibility constraint. This mechanism not only explains the emergence of the subharmonic mode but also selects the correct sign for $A$ leading to profiles actually observed in experiments, see Fig.~\ref{fig03-sup}. Obviously, this analysis does not, however, imply that an harmonic mode with a positive amplitude, $A>0$, is stable against subharmonic perturbations. Indeed, the above analysis is performed using, without loss of generality, an even function to describe the evolution of the wrinkled pattern ($y(\ell)=A\cos(q_0\ell) + B \cos(q_0 \ell/2)$). Having found the energetically favorable pattern in this case, we can use the translation invariance to generate equivalent patterns: $y(\ell-\pi/q_0)=-A\cos(q_0\ell) + B \sin(q_0 \ell/2)$. The sign of $A$ being now reversed, it implies that an harmonic mode with a positive amplitude is also unstable against subharmonic perturbations above the same threshold and leads to the same wrinkled pattern but translated. However, the amplitude of the subharmonic mode cannot be obtained with the linear equation (\ref{forcing-app2}) and is computed in Sec.~\ref{sec:num}.

\section{Numerical analysis of the main nonlinear equation}
\label{sec:num}

In order to confirm the period-doubling mechanism proposed in the main text, we analyze numerically the nonlinear equation (5) in the main text, namely
\begin{equation}
	\label{eqnonl}
	B_m y^{(IV)} + F y'' + K {\cal H}(y') + K_2 {\cal H}(y')^2=0.
\end{equation}
We use a rescaling similar to the one proposed in Sec.~\ref{app2} except for the function: $F=\bar{F} q_0^2 B_m$, $K=2 q_0^3 B_m$, $K_2=\bar{K}_2 q_0^3 B_m$, $q_0\ell = z$ and $y = u/(q_0\bar{K}_2)$. Equation (\ref{eqnonl}) reduces then to
\begin{equation}
	\label{eqnonl-univ}
	u^{(IV)} + \bar{F} u'' + 2{\cal H}(u') + {\cal H}(u')^2=0.
\end{equation}
We search solutions under the form
\begin{equation}
	\label{sum}
	u(z) = \sum_{k=1}^N c_k \cos(kz/2),
\end{equation}
with $c_k=2 q_0 (K_2/K) C_k$ where $C_k$ are the expansion coefficients introduced in the main text. Substituting the form (\ref{sum}) into (\ref{eqnonl-univ}) we obtain a system of $N$ equations with $N+1$ unknowns ($\bar{F}$ and $c_i$, $i=1,\ldots,N$). $N$ unknowns can be expressed in terms of $c_2$, for example. This last coefficient is determined from the inextensibility constraint.  

\begin{figure}[!hbtp]
\centerline{\includegraphics[width=6cm,clip]{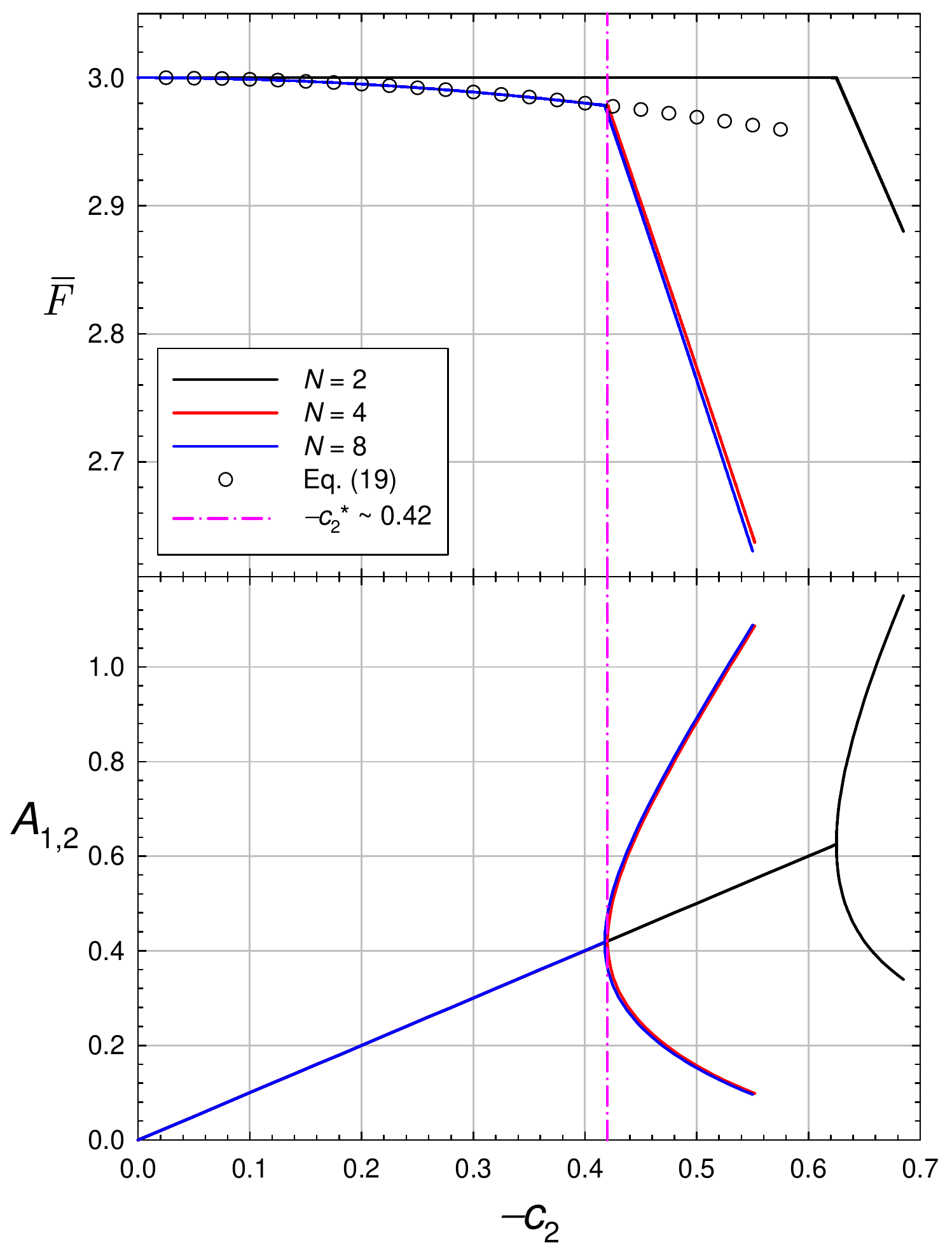}}
\caption{Evolution of ${\bar F}$ and the amplitudes $A_{1,2}$ (see Fig.~1 of the main text for the definition) as a function of the rescaled amplitude of the harmonic mode for several values of $N$.} 
\label{fig04-sup}
\end{figure}

For $N=2$, the computation is analytical and is already sufficient to understand the mechanism for the emergence of the subharmonic mode. Remembering that the terms $u$ and $u^2$ are operators that should be multiplied by the frequency of the modes on which they act, we obtain the following system of equations
\begin{eqnarray}
	c_2(3-\bar{F}) + \frac{c_1^2}{8}&=&0 \nonumber \\
	\frac{c_1}{16} \left(17-4\bar{F}+8 c_2 \right)&=&0.
\end{eqnarray}
This system admits two solutions. The first one is $\bar{F}=3$ and $c_1=0$ ($c_2$ being determined by the inextensibility constraint). This solution correspond to the evolution of the shape of the membrane without subharmonic mode. The second solution characterized by a subharmonic mode is
\begin{eqnarray}
	\label{sys-eq}
	\bar{F} &=& \frac{17}{4}+2 c_2 \\
	c_1^2&=& 8 c_2(\bar{F}-3)= 2 c_2\left(5+8 c_2 \right).
\end{eqnarray}
The expression for $\bar{F}$ is identical to Eq.~(\ref{eq-f}) obtained in Sec.~\ref{app2}. The amplitude of the harmonic mode, $c_2$, can be either positive or negative since the system, at the linear order, has an up-down symmetry. However, adding a quadratic nonlinearity breaks this symmetry, consequently the expression~(\ref{sys-eq}) for $\bar{F}$ is no longer invariant under a change of sign of $c_2$. For large enough (in absolute value) negative values of the amplitude $c_2$, {\it i.e.} $-c_2>5/8$, Eq.~(\ref{sys-eq}) shows that adding a subharmonic mode leads to a smaller value of $\bar{F}$ than the value obtained with the harmonic mode alone, {\it i.e.} $\bar{F}=3$. The situation is summarized in Figs.~2a and b in the main text.

\begin{figure}[!hbtp]
\centerline{\includegraphics[width=6cm,clip]{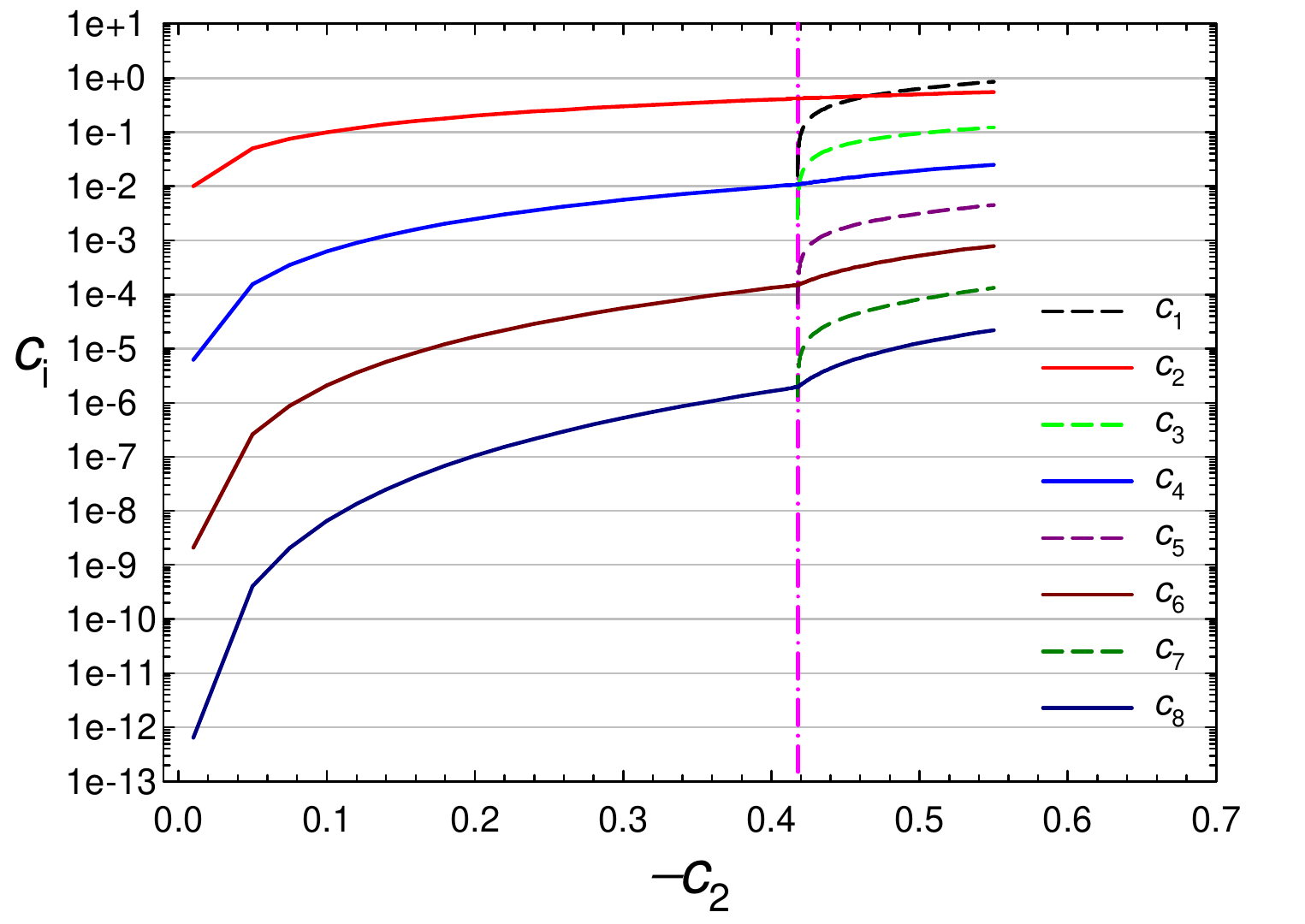}}
\caption{Evolution of the Fourier mode coefficients as a function of the rescaled amplitude of the harmonic mode for $N=8$.} 
\label{fig05-sup}
\end{figure}

For $N=4$, only the solution without subharmonic mode can still be obtained analytically in a simple form; it reads
\begin{eqnarray}
	\label{sol-n4}
	\bar{F} &=& 4-\sqrt{1+\left(\frac{c_2}{2}\right)^2}	\nonumber \\
	c_1&=&c_3 = 0 \nonumber \\
	2 c_4&=& 1-\sqrt{1+\left(\frac{c_2}{2}\right)^2}.
\end{eqnarray}
We see again that the quadratic nonlinearity breaks the up-down symmetry since for any sign of $c_2$, $c_4$ is always negative. The solution containing a subharmonic mode or solutions for larger $N$ are obtained numerically. In Fig.~\ref{fig04-sup}, we study the convergence for the evolution of ${\bar F}$ and $A_{1,2}$ (see Fig.~1 of the main text for the definition) when the solution is expanded as Eq.~(\ref{sum}) for several values of $N$. Convergence is essentially reached for $N=4$. In Fig.~\ref{fig05-sup}, we present the evolution of the coefficients of the Fourier modes as a function of the rescaled amplitude, $c_2$, of the harmonic mode for $N=8$. We find that the subharmonic mode emerges for $c_2=c_2^*\simeq -0.42$. Returning to the original variable, we find that the subharmonic mode emerges when the amplitude, $A$, of the harmonic mode reaches the value
\begin{equation}
	\label{a-tresh}
	\frac{A}{\lambda_0}=-\frac{0.42}{4\pi (K_2/K)}.
\end{equation}

\begin{figure}[!hbtp]
\centerline{\includegraphics[width=6cm,clip]{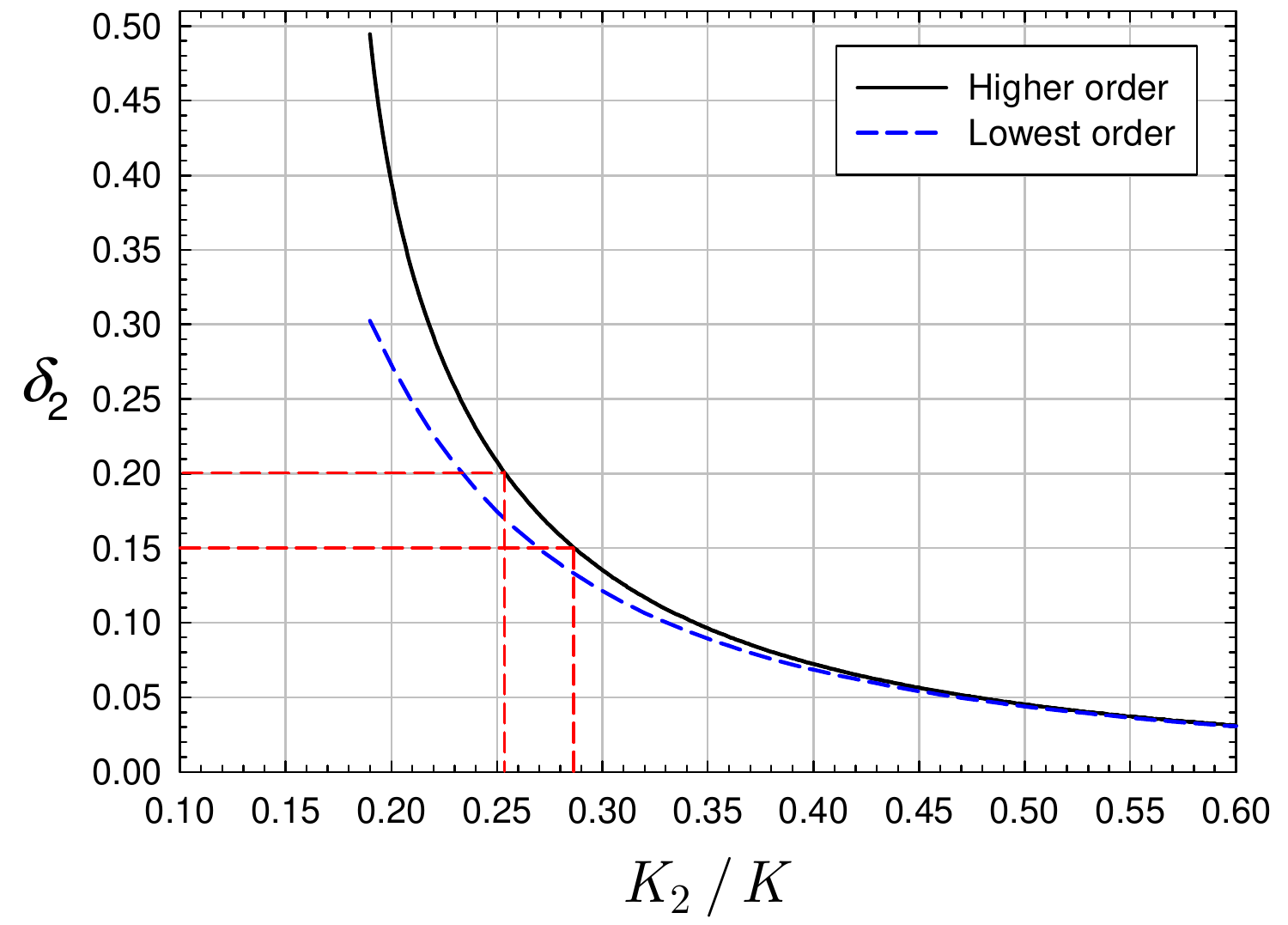}}
\caption{Threshold, $\delta_2$, of the period-doubling instability as a function of the strength of the nonlinearity $K_2/K$. The low-order relation (\ref{delta2-lowest}) (blue dashed curved) is compared to an higher order relation obtained from Eqs.~(\ref{a-tresh}) and (\ref{a-high}) (black solid curve).} 
\label{fig06-sup}
\end{figure}

\begin{figure}[!hbtp]
\centerline{\includegraphics[width=6cm,clip]{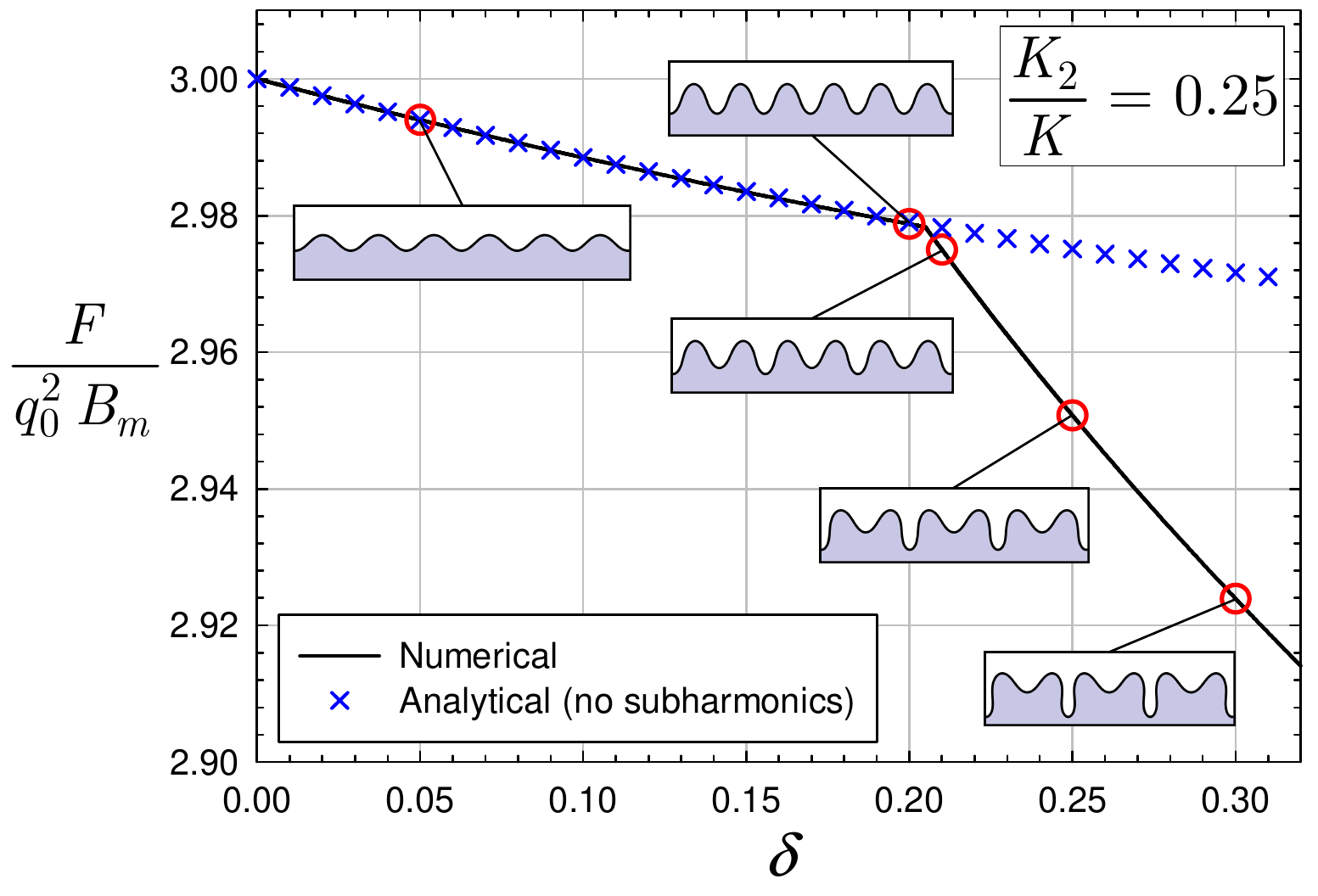}}
\caption{Numerical evolution of $F$ as a function of the relative compression $\delta$. The analytical expression (\ref{sol-n4}) for the evolution of $F$ without subharmonic mode is also presented.} 
\label{fig07-sup}
\end{figure}

The amplitude of the harmonic mode is related to the relative compression $\delta$ through the inextensibility constraint. We thus obtain an expression for the threshold in compression, $\delta_2$, for the emergence of the subharmonic mode. At the lowest order, we have $A/\lambda_0 = \sqrt{\delta}/\pi$ which leads to 
\begin{equation}
	\label{delta2-lowest}
	\delta_2 \simeq \left(\frac{0.105}{(K_2/K)}\right)^2.
\end{equation}
An higher order expression $A/\lambda_0$ as a function of $\delta$ reads
\begin{equation}
	\label{a-high}
\frac{A}{\lambda_0}=\frac{\sqrt{\delta}}{\pi}\left(1-\frac{3}{8}\delta-\frac{17}{128}\delta^2\right).
\end{equation}
This last expression together with (\ref{a-tresh}) yield an higher order relation for $\delta_2$ which is plotted, together with (\ref{delta2-lowest}), in Fig.~\ref{fig06-sup}. 

Finally, in Fig.~\ref{fig07-sup}, we present the numerical evolution of the pressure $F$ as a function of the relative compression $\delta$ together with corresponding profiles of the membrane.
\\

\begin{center}
\line(1,0){250}
\end{center}

[1] Nase, J., Lindner, A. \& Creton, C. Pattern formation during deformation of a confined viscoelastic layer: From a viscous liquid to a soft elastic solid, \textit{Physical Review Letters} {\bf101}, 074503 (2008). \\

[2] Efimenko, K., Rackaitis, M., Manias, E., Vaziri, A., Mahadevan, L., Genzer, J. Nested self-similar wrinkling patterns in skins, \textit{Nature Materials} {\bf4}, 293-297 (2005).\\

[3] Landau, L. D., \& Lifshitz, E. M. {\it Theory of Elasticity}, Pergamon, NY, 3rd edn, 1986.


\begin{thebibliography}{99}
\bibitem{witt07} Witten, T. A. Stress focusing in elastic sheets. {\it Rev. Mod. Phys.} {\bf 79}, 643-675 (2007).
\bibitem{bowd98} Bowden, N., Brittain, S., Evans, A. G., Hutchinson, J. W. \& G. M. Whitesides. Spontaneous formation of ordered structures in thin films of metals supported on an elastomeric polymer. {\it Nature} {\bf 393}, 146-149 (1998).
\bibitem{cerd03} Cerda, E. \& Mahadevan, L. Geometry and physics of wrinkling. {\it Phys. Rev. Lett.} {\bf 90}, 074302 (2003).
\bibitem{vandeparre07} Vandeparre, H. {\it et al.} Slippery or Sticky Boundary Conditions: Control of Wrinkling in Metal-Capped Thin Polymer Films by Selective Adhesion to Substrates. {\it Phys. Rev. Lett.} {\bf99}, 188302 (2007).
\bibitem{vandeparre08} Vandeparre, H. \& Damman, P. Wrinkling of Stimuloresponsive Surfaces: Mechanical Instability Coupled to Diffusion. {\it Phys. Rev. Lett.} {\bf101}, 124301 (2008).
\bibitem{huan07} Huang, J. {\it et al.} Capillary wrinkling of floating thin polymer films. {\it Science} {\bf 317}, 650-653 (2007).
\bibitem{jian07} Jiang, H. {\it et al.} Finite deformation mechanics in buckled thin films
on compliant supports. {\it PNAS} {\bf 104}, 15607-15612 (2007).
\bibitem{poci08} Pocivavsek, L.  {\it et al.} Stress and fold localization in thin elastic membranes. {\it Science} {\bf 320}, 912-916 (2008).
\bibitem{efim04} Efimenko, K. {\it et al.} Nested self-similar wrinkling patterns in skins. {\it Nature Mater.} {\bf 4}, 293-297 (2005).  
\bibitem{diam01} Diamant, H., Witten, T. A., Ege, C., Gopal, A. \& Lee, K. Y. C. Topography and instability of monolayers near domain boundaries. {\it Phys. Rev. E} {\bf 63}, 061602 (2001).
\bibitem{strupler} Strupler, M. \textit{et al.}  Second harmonic microscopy to quantify renal interstitial Þbrosis and arterial remodeling. {\it J. Biomed. Opt.} {\bf13}, 054041 (2008)
\bibitem{rich75} Richman, D.~P., Stewart, R.~M., Hutchinson, J.~W. \& Caviness, V.~S., Jr. Mechanical model of brain convolutional development. {\it Science}, {\bf 189}, 18-21 (1975).
\bibitem{toro05} Toro, R. \& Burnod, Y. A Morphogenetic Model for the Development of Cortical Convolutions. {\it Cereb. Cortex} {\bf 15}, 1900-1913 (2005).
\bibitem{kuck04} K\"ucken, M. \& Newell, A.~C. A model for fingerprint formation. {\it Europhys. Lett.} {\bf 68}, 141-146 (2004).
\bibitem{feig78} Feigenbaum, M. J. Quantitative universality for a class of nonlinear transformations. {\it J. Stat. Phys.} {\bf 19}, 25-52 (1978).
\bibitem{feig79} Feigenbaum, M. J. The universal metric properties of nonlinear transformations. {\it J. Stat. Phys.} {\bf 21}, 669-706 (1979).
\bibitem{libc82} Libchaber, A., Laroche, C. \& Fauve, S. Period doubling cascade in mercury, a quantitative measurement. {\it J. Physique} {\bf 43}, L211-L216 (1982).
\bibitem{guev81} Guevara, M. R., Glass, L. \& Shrier, A. Phase locking, period-doubling bifurcations, and irregular dynamics in periodically stimulated cardiac cells. {\it Science} {\bf 214}, 1350-1353 (1981). 
\bibitem{fox02} Fox, J. J., Bodenschatz, E. \& Gilmour, R. F. Period-doubling instability and memory in cardiac tissue. {\it Phys. Rev. Lett.} {\bf 89}, 138101 (2002).
\bibitem{berg07} Berger, C. M. {\it et al.} Period-doubling bifurcation to alternans in paced-cardiac tissue: Crossover from smooth to border-collision characteristics. {\it Phys. Rev. Lett.} {\bf 99} 058101 (2007).
\bibitem{melo95} Melo, F., Umbanhowar, P. B. \& Swinney, H. L. Hexagons, kinks, and disorder in oscillated granular layers. {\it Phys. Rev. Lett.} {\bf 75}, 3838-3841 (1995).
\bibitem{venk98} Venkataramani, S. C. \& Ott, E. Spatiotemporal bifurcation phenomena with temporal period doubling: patterns in vibrated sand. {\it Phys. Rev. Lett.} {\bf 80}, 3495-3498 (1998).
\bibitem{gile09} Gilet, T. \& Bush, J. Chaotic bouncing of a droplet on a soap film. {\it Phys. Rev. Lett.} {\bf 102}, 014501 (2009).
\bibitem{lose96} Losert, W., Shi, B. Q. \& Cummins, H. Z. Spatial period-doubling instability of dendritic arrays in directional solidification. {\it Phys. Rev. Lett.} {\bf 77}, 889-891 (1996).
\bibitem{koit70} Hutchinson, J. W. \& Koiter, W. T. Postbuckling theory. {\it Appl. Mech. Rev.} {\bf 23}, 1353-1366  (1970).
\bibitem{groe01} Groenewold, J. Wrinkling of plates coupled with soft elastic media. {\it Physica A} {\bf 298}, 32-45 (2001).
\bibitem{mcla62} McLachlan, N. W. {\it Theory and application of Mathieu functions}, Dover, New-York, 1962.
\bibitem{blan72} Blanch, G., Chapter 20: Mathieu Functions, in {\it Handbook of Mathematical Functions with Formulas, Graphs, and Mathematical Tables}, Eds. Abramowitz, M. \& Stegun, I. A., Dover, New-York, 1972.
\bibitem{sanm84} Sanmart\'in, J. R. O Botafumeiro: Parametric pumping in the Middle Ages. {\it Am. J. Phys.} {\bf 52}, 937-945 (1984).
\bibitem{vand00} Van den Broeck, C. \& Bena I. Parametric Resonance Revisited, in ``Stochastic Processes in Physics, Chemistry, and Biology'', Eds. Freund, J. A. \& P\"oschel T. {\it Lect. Notes Phys.} {\bf 557}, 257-267 (2000).
\end{thebibliography}
\end{document}